\documentclass{SciPost}

\hypersetup{
    colorlinks,
    linkcolor={red!50!black},
    citecolor={blue!50!black},
    urlcolor={blue!80!black}
}

\usepackage[bitstream-charter]{mathdesign}
\DeclareSymbolFont{usualmathcal}{OMS}{cmsy}{m}{n}
\DeclareSymbolFontAlphabet{\mathcal}{usualmathcal}

\fancypagestyle{SPstyle}{
\fancyhf{}
\lhead{\colorbox{scipostblue}{\bf \color{white} ~SciPost Physics Proceedings }}
\rhead{{\bf \color{scipostdeepblue} ~Submission }}

\fancyfoot[C]{\textbf{\thepage}}
}

\begin{document}

\pagestyle{SPstyle}

\begin{center}{\Large \textbf{\color{scipostdeepblue}{
Probes of flavour symmetry and violation
with top quarks in ATLAS and CMS\\
}}}\end{center}

\begin{center}\textbf{
Miriam Watson\textsuperscript{1$\star$}, on behalf of the ATLAS and CMS Collaborations
}\end{center}

\begin{center}
{\bf 1} University of Birmingham, UK
\\[\baselineskip]
$\star$ \href{mailto:Miriam.Watson@cern.ch}{\small Miriam.Watson@cern.ch}
\end{center}

\definecolor{palegray}{gray}{0.95}
\begin{center}
\colorbox{palegray}{
  \begin{tabular}{rr}
  \begin{minipage}{0.36\textwidth}
    \includegraphics[width=60mm,height=1.5cm]{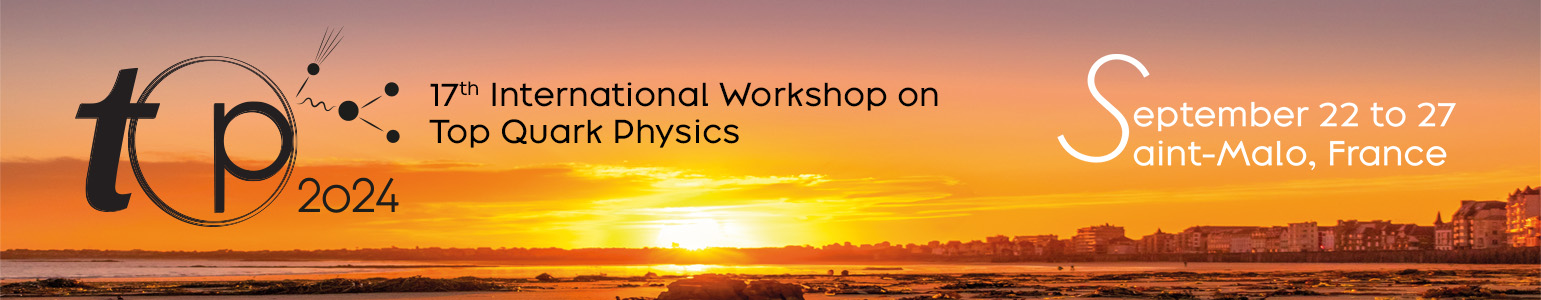}
  \end{minipage}
  &
  \begin{minipage}{0.55\textwidth}
    \begin{center} \hspace{5pt}
    {\it The 17th International Workshop on\\ Top Quark Physics (TOP2024)} \\
    {\it Saint-Malo, France, 22-27 September 2024
    }
    \doi{10.21468/SciPostPhysProc.?}\\
    \end{center}
  \end{minipage}
\end{tabular}
}
\end{center}

\section*{\color{scipostdeepblue}{Abstract}}
\textbf{\boldmath{%
    Results are presented of searches and measurements in the top quark sector by the ATLAS and CMS experiments.  These analyses use data from proton-proton collisions at a centre-of-mass energy of $13$~TeV, recorded during Run 2 at the Large Hadron Collider and corresponding to integrated luminosities of 138--140 fb$^{-1}$.  Searches are carried out for charged lepton flavour violation, baryon number violation and the presence of neutral heavy leptons.  A precise measurement of lepton flavour universality between electrons and muons originating from top quark-antiquark events is also presented.
}}

\vspace{\baselineskip}

\noindent\textcolor{white!90!black}{%
\fbox{\parbox{0.975\linewidth}{%
\textcolor{white!40!black}{\begin{tabular}{lr}%
  \begin{minipage}{0.6\textwidth}%
    {\small Copyright [2025] CERN for the benefit of the ATLAS and CMS Collaborations. CC-BY-4.0 license.
      \newline
    This work is a submission to SciPost Phys. Proc. \newline
    License information to appear upon publication. \newline
    Publication information to appear upon publication.}
  \end{minipage} & \begin{minipage}{0.4\textwidth}
    {\small \mbox{} \newline \mbox{} \newline
      Received Date \newline Accepted Date \newline Published Date}%
  \end{minipage}
\end{tabular}}
}}
}




\section{Introduction}
\label{sec:intro}
As the heaviest fundamental particle in the Standard Model (SM), the top quark has a unique role in validating the predictions of the SM and in searches for new physics. 
The large data samples accumulated by the ATLAS~\cite{ATLAS} and CMS~\cite{CMS} Collaborations during the Run 2 data-taking period of the Large Hadron Collider (LHC)~\cite{LHC} allow stringent tests of the top quark sector in the SM.  These include tests of the universality of lepton couplings, and searches for possible deviations from the SM predictions via violations of charged-lepton flavour, baryon number or lepton number.  Six new analyses are described here.  These are based on the full Run 2 $pp$ data samples from ATLAS or CMS, at a centre-of-mass energy of $13$~TeV\@.  The ATLAS analyses use an integrated luminosity of 140 fb$^{-1}$ collected in 2015--2018, while those from CMS use 138 fb$^{-1}$ collected during 2016--2018.  

\section{Lepton Flavour Universality}
\label{sec:LFU}

The assumption of lepton flavour universality (LFU) is a fundamental axiom of the SM, \textit{i.e.} the couplings of the charged leptons to the $W$ or $Z$ bosons should be independent of the lepton flavour.  This assumption can be tested at high momentum using $W$ bosons originating from top quark decays. The ATLAS analysis~\cite{ATLAS:2024tlf} tests LFU between electrons and muons in the dileptonic decays of $t\bar{t}$ events. It exploits the similar kinematics of $Z\rightarrow\ell^+\ell^-$ decays to reduce the impact of lepton identification uncertainties, through the measurement of the double ratio
\begin{equation}
 \label{eq:ratio}
R^{\mu/e}_{WZ} = \frac{R^{\mu/e}_{W}}{\sqrt{R^{\mu\mu/ee}_{Z}}} = \frac{{\cal B}(W\rightarrow\mu\nu)}{{\cal B}(W\rightarrow e\nu)}\cdot \sqrt{\frac{{\cal B}(Z\rightarrow ee)}{{\cal B}(Z\rightarrow \mu\mu)}}\, ,
\end{equation}
where the $t\bar{t}$ events are selected in the $ee$, $e\mu$ and $\mu\mu$ channels with either 1 or 2 $b$-tagged jets, and the $Z$ boson samples are selected inclusively in the $ee$ or $\mu\mu$ channels.  Reweighting factors are applied to the muons as a function of transverse momentum $p_{\mathrm{T}}$ and absolute pseudorapidity $|\eta|$ to reduce the kinematic differences between the selected electrons and muons.  The lepton isolation efficiencies are measured in-situ as a function of $p_{\mathrm{T}}$ and $|\eta|$, separately for electons and muons in the $Z$ and $t\bar{t}$ samples; this allows the different environments in the two types of events to be taken into account.

\begin{figure}[htb]
  \begin{centering}
    \hspace*{-12mm}
  \parbox{.54\textwidth}{\includegraphics[width=.55\textwidth]{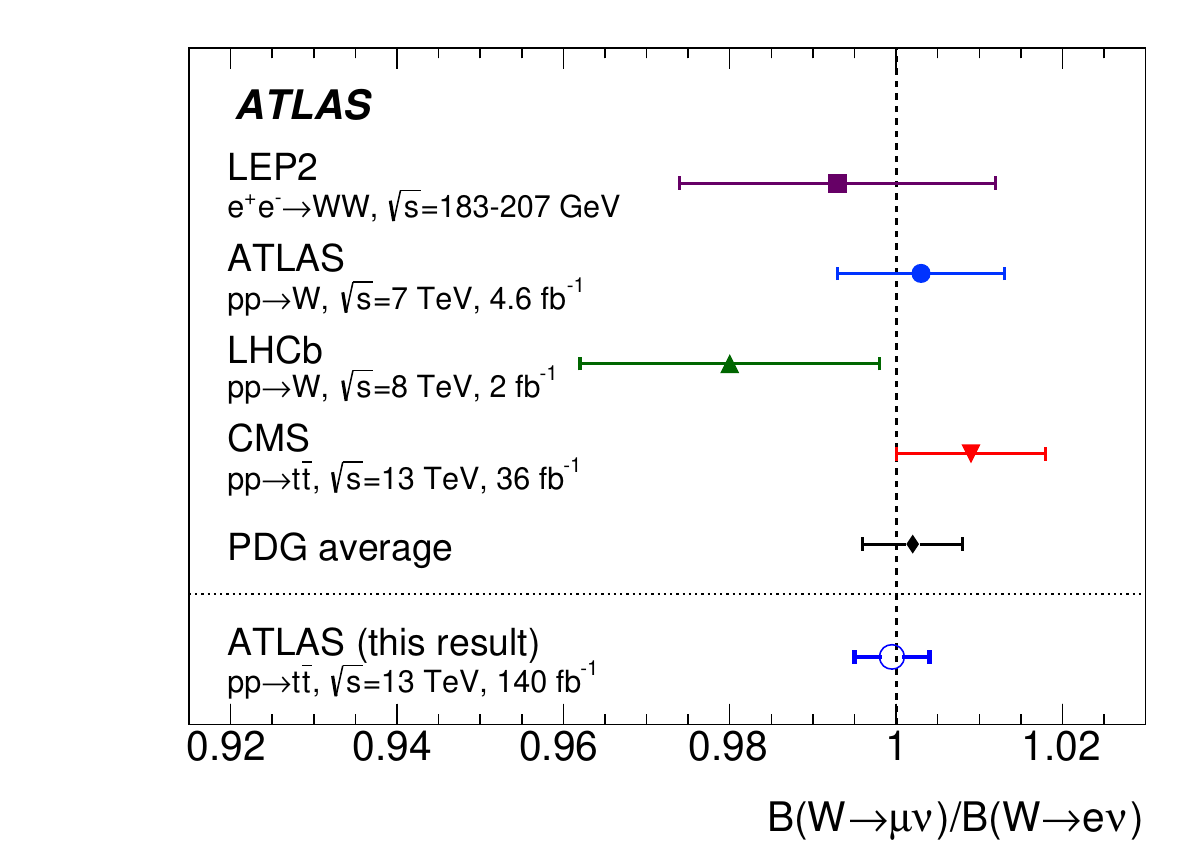}\vspace{-7mm}\center{(a)}}
    \hspace*{6mm}
\parbox{.44\textwidth}{\includegraphics[width=.45\textwidth]{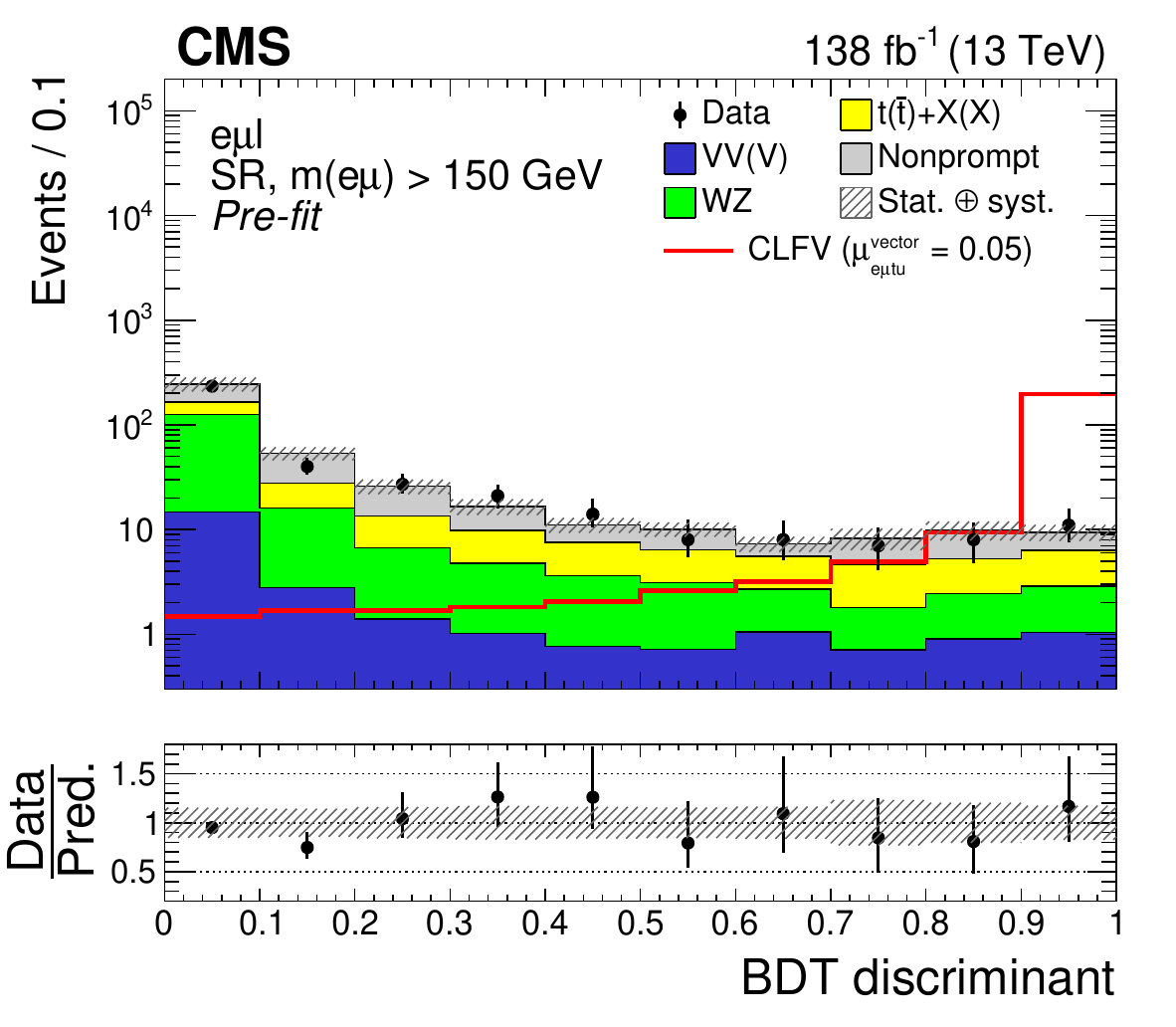}\vspace{-5mm}\center{(b)}}
 \caption{\label{fig:LFU_cLFV}(a) Summary of ${\cal B}(W\rightarrow\mu\nu)/{\cal B}(W\rightarrow e\nu)$ measurements, including the ATLAS result~\cite{ATLAS:2024tlf}; (b) Distribution of the BDT discriminant in the $m(e\mu)>150$~GeV region for the CMS $e\mu$ trilepton cLFV analysis~\cite{arXiv:2312.03199}.}
 \end{centering}
 \end{figure}

The branching fractions are extracted using a simultaneous likelihood fit to the $e\mu$ yields in the $t\bar{t}$ events, the $ee$ and $\mu\mu$ yields in the $Z$ events, and the $\ell\ell$ ($\ell = e,\mu$) invariant mass distribution in the same-flavour $t\bar{t}$ samples.  They are parametrised as deviations from the SM branching fractions.  The relative $W$ branching fractions are extracted by applying the precise external $Z$ branching fraction ratio from LEP and SLD~\cite{LEP_SLD} to the fitted value in Equation~\ref{eq:ratio}, leading to
\begin{equation}
 \label{eq:result}
 R^{\mu/e}_{W} = \frac{{\cal B}(W\rightarrow\mu\nu)}{{\cal B}(W\rightarrow e\nu)} = 0.9995\pm0.0045\, ,
\end{equation}
where the dominant uncertainties arise from PDF, modelling and lepton uncertainties.  This measurement is consistent with the assumption of LFU.  It is the most precise test of $e$-$\mu$ universality to date and improves on the previous PDG average~\cite{PDG:2022}, as shown in Figure~\ref{fig:LFU_cLFV}(a).

\section{Charged Lepton Flavour Violation}
\label{sec:cLFV}
Within the SM, charged lepton flavour violation (cLFV) is expected due to neutrino oscillations, with an extremely small branching ratio of ${\cal O}(10^{-55})$~\cite{Calibbi}.  An observation of cLFV in the LHC data would therefore provide a strong indication of beyond-the-Standard-Model (BSM) physics, with possible contributions from leptoquarks, supersymmetry or two-Higgs-doublet models.  Recent results from CMS and ATLAS~\cite{arXiv:2312.03199,ATLAS:2024njy,CMS:PAS22.011} all consider cLFV processes in single-top production (``production'') or in $t\bar{t}$ decay (``decay''). Limits on a possible cLFV contribution are extracted in a model-independent Effective Field Theory (EFT) approach, in which it is assumed that the mass scale of new physics $\Lambda$ is significantly larger than the LHC energy scale.  This allows limits to be placed on Wilson coefficients $|C|$ for vector, scalar and tensor $tq\ell\ell^{\prime}$ dimension-6 operators, where $q$ is an up- or a charm-quark. Table~\ref{tab:cLFVsummary} summarises the three analyses to be discussed in Sections~\ref{sec:CMSemu} to \ref{sec:CMSmutau}.

\begin{table}[htb]
\caption{\label{tab:cLFVsummary} Summary of the cLFV analysis requirements.}
\centering
\begin{tabular}{|l|l|l|l|}\hline
  Analysis & CMS $e\mu$ trilepton 
  & ATLAS $\mu\tau$ trilepton 
  & CMS $\mu\tau$ hadronic 
  \\
\hline
cLFV vertex  & $te\mu q~(q=u,c)$ & $t\mu\tau q~(q=u,c)$ & $t\mu\tau q~(q=u,c)$ \\
cLFV leptons & $e^\pm\mu^\mp$ & $\mu^\pm\tau^\mp$ & $\mu^\pm\tau^\mp$ \\
SM W decay   & $e^\pm$ or $\mu^\pm$ (+$\nu$) & $\mu^\pm$ (+$\nu$) & $q \bar{q}^\prime$ \\
Signature    & $e^\pm\mu^\mp\ell$ & $\mu^\pm\mu^\pm\tau_{\mathrm had}^\mp$ & $\mu^\pm\tau_{\mathrm had}^\mp$+jets \\
No. of $b$-tagged (all) jets & 1 (1 or 2) & 1 ($\geq 1$) & 1 ($\geq 3$) \\
\hline
\end{tabular}
\end{table}

\subsection{cLFV in the $e\mu$ Trilepton Channel}
\label{sec:CMSemu}
The CMS $e\mu$ trilepton analysis~\cite{arXiv:2312.03199} makes use of Boosted Decision Tree (BDT) discriminants to separate the background from the cLFV signal.  The BDT is trained in two regions, above and below an invariant mass $m(e\mu)$ of 150 GeV; the higher mass region is enriched in the production process and is shown in Figure~\ref{fig:LFU_cLFV}(b), while the lower masses target $t\bar{t}$ decays.  Each of the EFT Wilson coefficients is extracted separately in binned likelihood fits, combining the production and decay processes in each case. No significant excess is observed over the SM expectation, allowing $95\%$ CL upper limits to be extracted on the Wilson coefficients $C$. These are also converted to branching fraction limits for ${\cal B}(t\rightarrow q\ell\ell^\prime)$.  The results are summarised in the third column of Table~\ref{tab:cLFVlimits}.

\begin{table}[htb]
\caption{\label{tab:cLFVlimits} Limits on Wilson coefficients $|C|$ and branching ratios ${\cal B}$ at 95\% CL. The range of limits corresponds to the different operator types: the limits on $|C|$ vary from smallest to largest for tensor, vector and scalar operators, respectively, while the relative ordering of the limits is reversed for ${\cal B}$.}
\centering
\begin{tabular}{|l|c|c|c|c|}\hline
  Analysis & $q$ & CMS $e\mu$ trilepton 
  & ATLAS $\mu\tau$ trilepton 
  & CMS $\mu\tau$ hadronic 
  \\
\hline
cLFV vertex  &    & $te\mu q~(q=u,c)$ & $t\mu\tau q~(q=u,c)$ & $t\mu\tau q~(q=u,c)$ \\
$|C|/\Lambda^2$ [GeV$^{-2}$] & $u$ & 0.02--0.10 & 0.10--0.44 & 0.045--0.18 \\
 & $c$ & 0.09--0.42 & 0.36--1.8 & 0.19--0.82 \\
${\cal B}(t\rightarrow q\ell\ell^\prime)$ [$10^{-6}$] & $u$ & 0.012--0.032 & 0.20-0.52 & 0.04--0.12 \\
 & $c$ & 0.22--0.50 & 3.4--6.7 & 0.81--2.05 \\
\hline
\end{tabular}
\end{table}

\subsection{cLFV in the $\mu\tau$ Trilepton Channel}
\label{sec:ATLASmutau}
In the ATLAS $\mu\tau$ cLFV analysis~\cite{ATLAS:2024njy}, the $\tau$ leptons are selected in their hadronic decay modes, $\tau_{\mathrm had}$.  The discriminating variable is $H_{\mathrm T}$, calculated from the scalar sum of the transverse momenta of jets and leptons in the event; the cLFV signal for the production channel tends to lie at high $H_{\mathrm T}$. No significant excess of events is found in the data, leading to the limits shown in the fourth column of Table~\ref{tab:cLFVlimits}; these improve on the previous indirect limits by factors of 7--41 and the analysis is statistically limited.

\begin{figure}[htb]
\parbox{.54\textwidth}{\includegraphics[width=.55\textwidth]{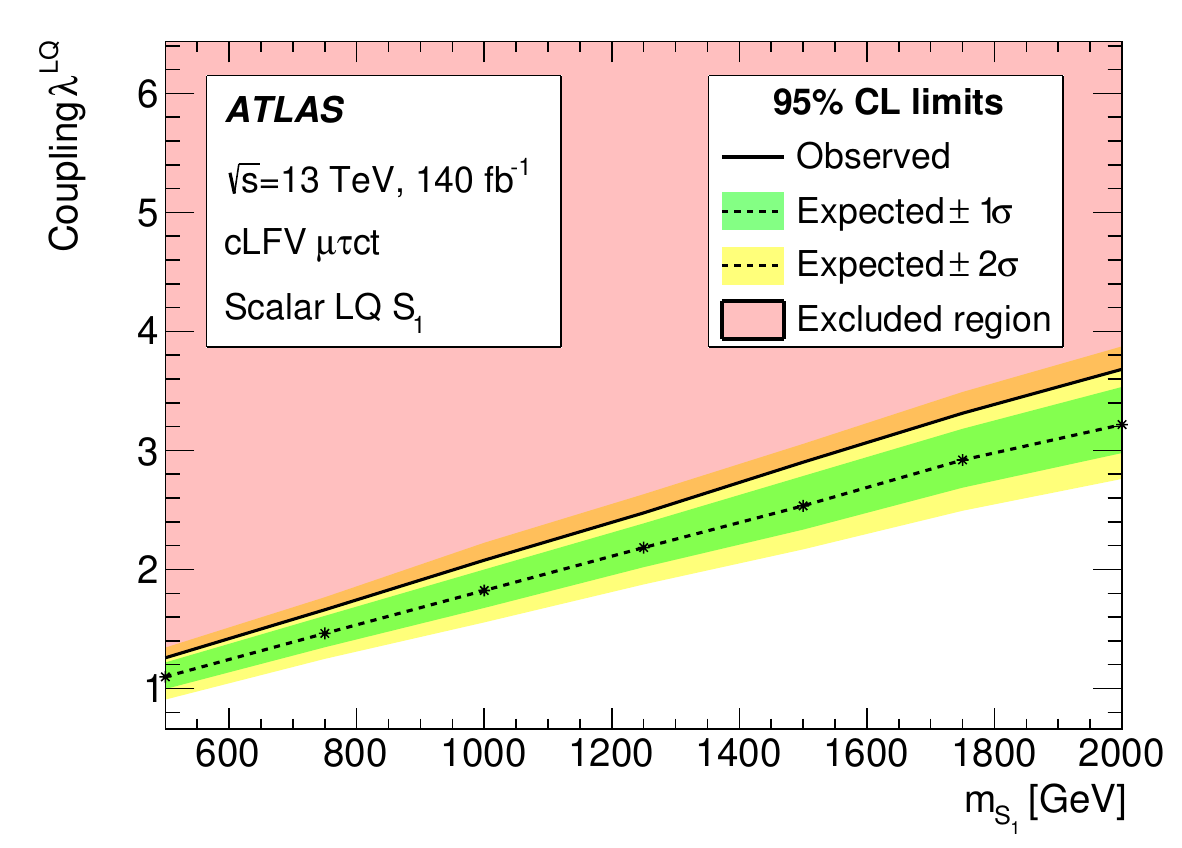}\vspace{-4mm}\center{(a)}}
\parbox{.44\textwidth}{\includegraphics[width=.45\textwidth]{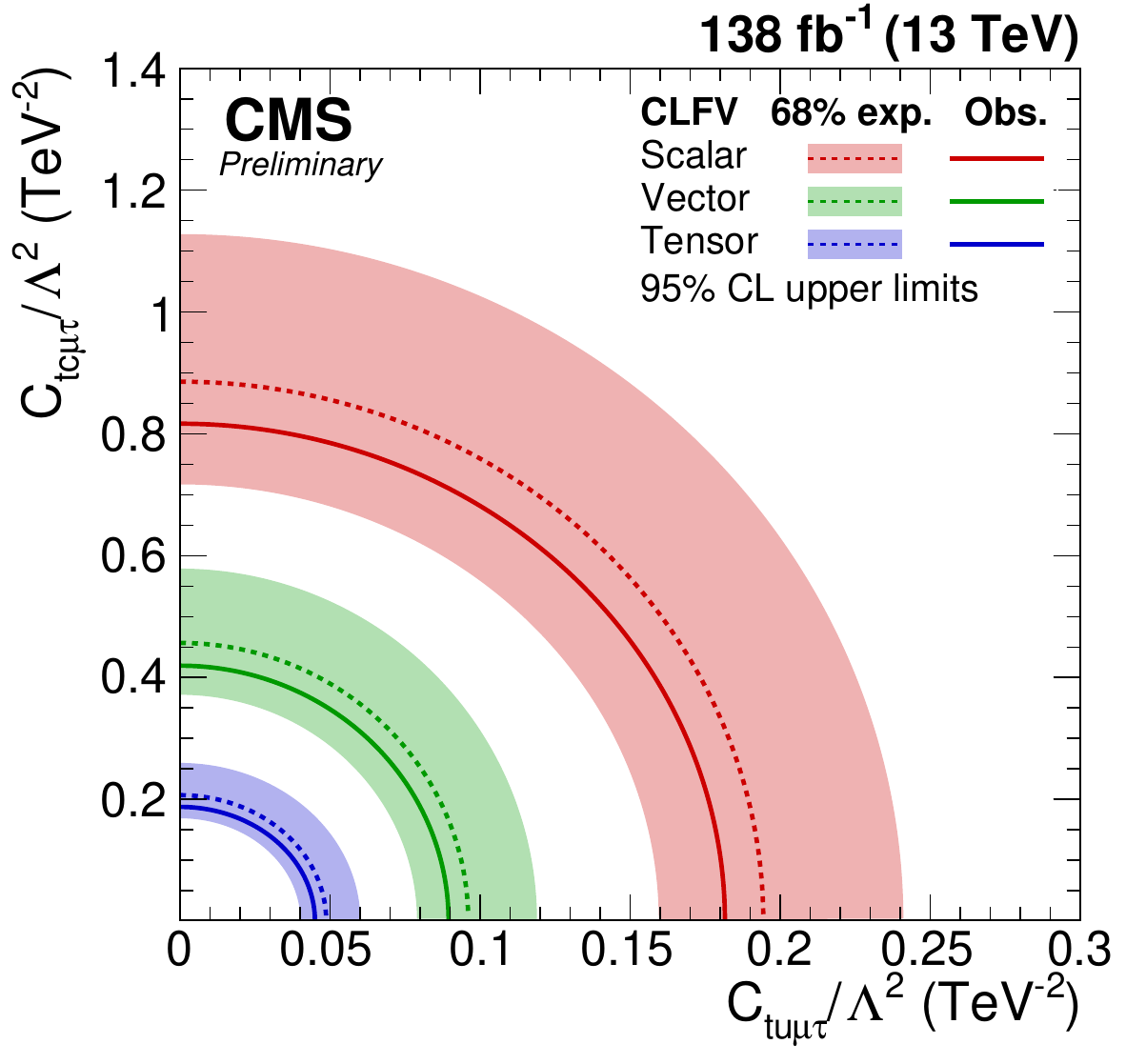}\vspace{-7mm}\center{(b)}}
 \caption{\label{fig:cLFV}(a) 95\% CL upper limits as a function of the leptoquark coupling strength $\lambda_{\mathrm LQ}$ and mass $m_{\mathrm S_1}$ in the ATLAS $\mu\tau$ cLFV analysis~\cite{ATLAS:2024njy}; (b) Exclusion contours for Wilson coefficients relating to scalar, vector and tensor Lorentz structures in the CMS $\mu\tau$ cLFV note~\cite{CMS:PAS22.011}.}
\end{figure}

The same data are used to search for a scalar leptoquark $S_1$~\cite{Dorsner}, assuming fixed hierarchical couplings which reduce by a factor of 10 for each generation of quarks or leptons.  The highest coupling value is for the $t$-$\tau$ coupling, $\lambda_{\mathrm{t}\tau}$, expressed in terms of a universal coupling strength $\lambda_{\mathrm LQ}$ ($=\lambda_{\mathrm{t}\tau}/10$).  Limits are extracted on $\lambda_{\mathrm LQ}$ and on the leptoquark mass $m_{\mathrm S_1}$.  Upper limits are set from $\lambda_{\mathrm LQ}=1.3$ to $\lambda_{\mathrm LQ}=3.7$ for masses in the range $m_{\mathrm S_1}=0.5$~TeV to $m_{\mathrm S_1}=2.0$~TeV, at 95\% CL, as illustrated in Figure~\ref{fig:cLFV}(a).

\subsection{cLFV in the $\mu\tau$ Hadronic Channel}
\label{sec:CMSmutau}
The latest CMS cLFV analysis~\cite{CMS:PAS22.011} considers the $\mu\tau$ channel, in which the $W$ boson from the SM top quark decays hadronically and the $\tau$ leptons are also identified in their hadronic decay modes.  Kinematic reconstruction of the $W$ and $t$ decay using a $\chi^2$ minimisation improves the separation of the signal from the dileptonic $t\bar{t}$ background.  A multiclass deep neural network (DNN) is trained to classify three nodes: the cLFV signals from production or decay, and the $t\bar{t}$ background.  This leads to an overall DNN score based on 28 kinematic variables.  The data are consistent with the SM expectation, leading to an improvement of a factor of two relative to the previous Wilson coefficient limits for the $\mu\tau$ channel.  These limits are shown in the final column of Table~\ref{tab:cLFVlimits}, together with the branching ratios, and are shown in Figure~\ref{fig:cLFV}(b) for the separate scalar, vector and tensor contributions.

\section{Baryon Number Violation}
\label{sec:BNV}

\begin{figure}[htb]
\parbox{.39\textwidth}{\includegraphics[width=.4\textwidth]{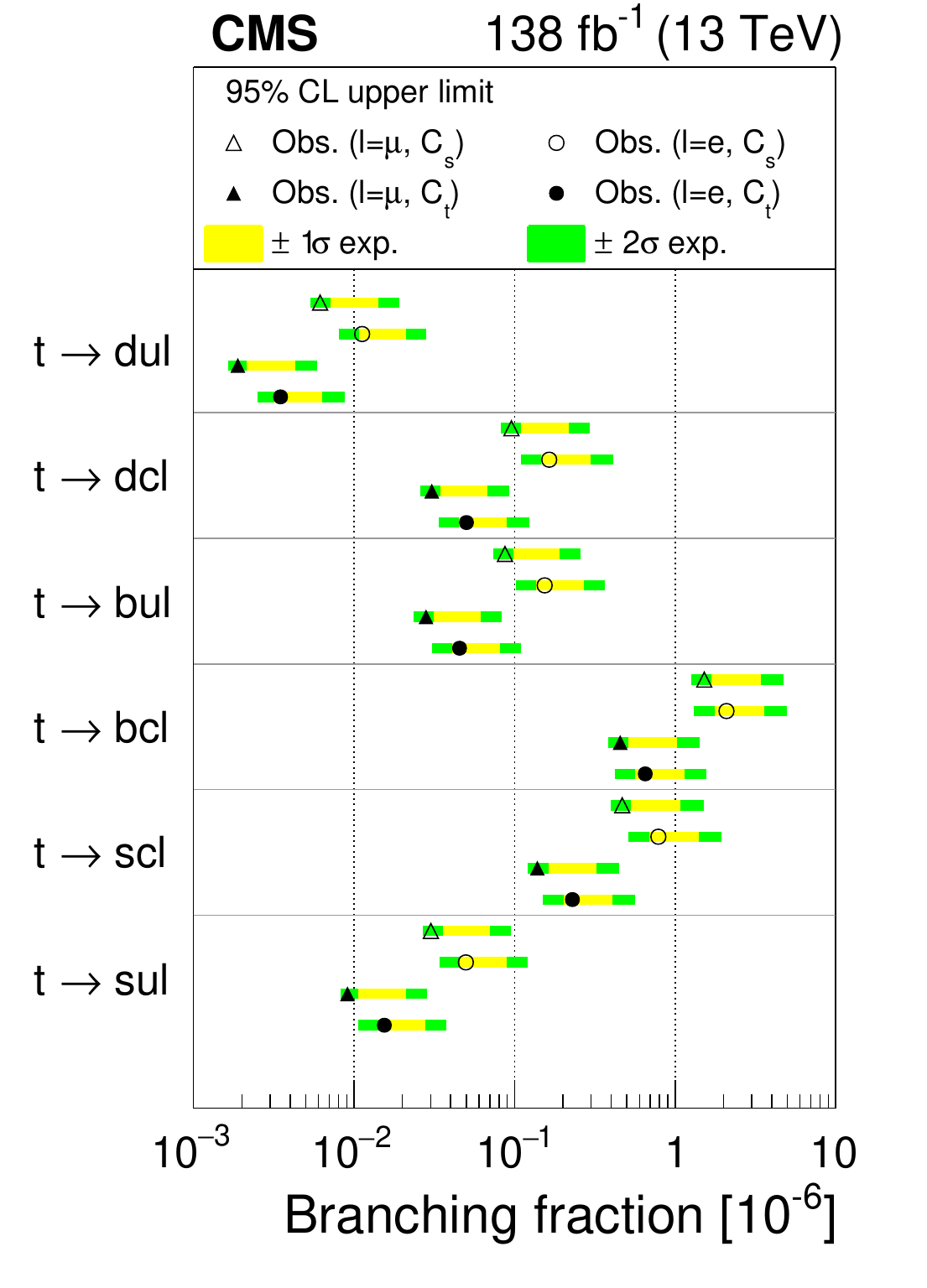}\vspace{-7mm}\center{(a)}}
\parbox{.59\textwidth}{\includegraphics[width=.6\textwidth]{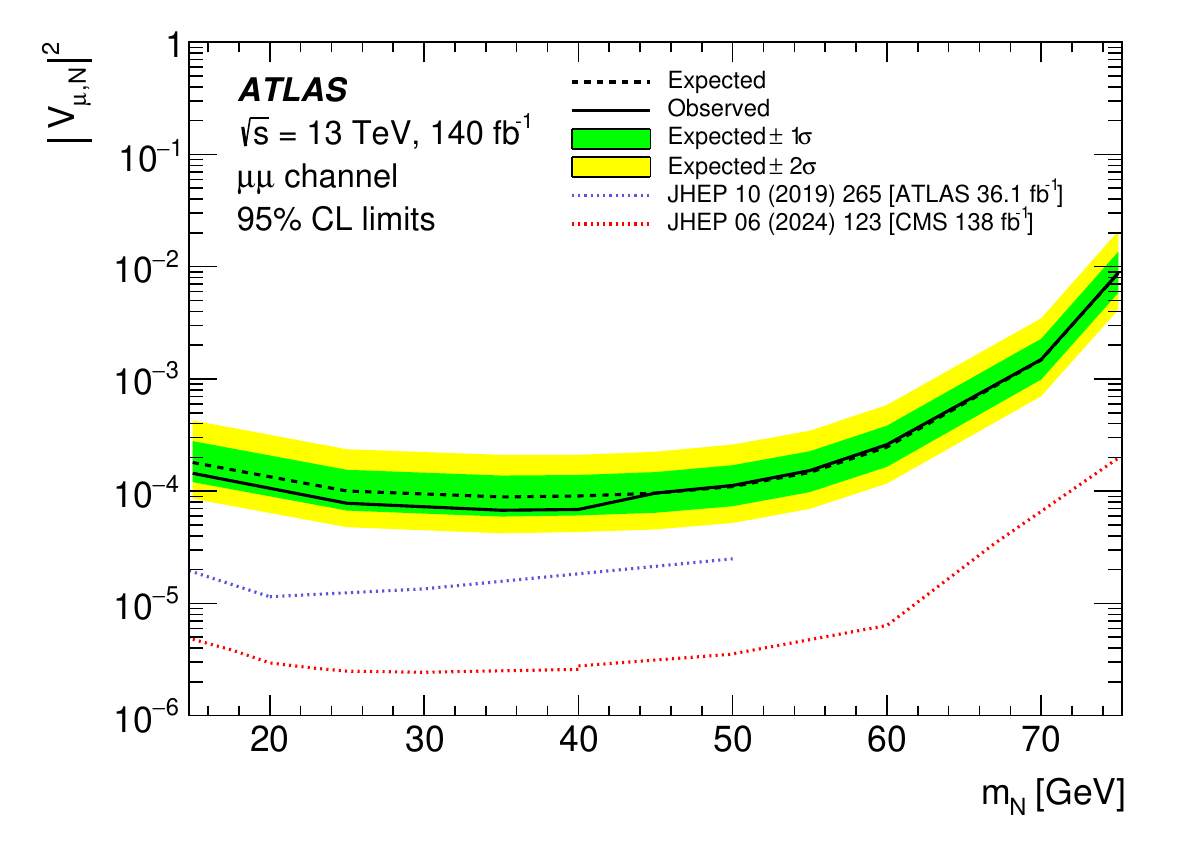}\vspace{-7mm}\center{(b)}}
 \caption{\label{fig:BNV_HNL}(a) Upper limits on the branching fractions of the top quark BNV decays in the CMS experiment~\cite{CMS:2024dzv}; (b) Upper limits on the strength of HNL mixing with muon neutrinos, including the ATLAS $t\bar{t}$ result~\cite{ATLAS:2024fcs}.}
\end{figure}

Baryon number is conserved in the SM but does not correspond to a fundamental symmetry. It can be violated by small non-perturbative effects, and baryon number violation (BNV) can be enhanced in BSM processes.  The CMS BNV analysis~\cite{CMS:2024dzv} is similar in structure to the cLFV analyses in Section~\ref{sec:cLFV} and includes the single-top production and $t\bar{t}$ decay processes; however, the effective interaction is now a $tqq^{\prime}\ell$ vertex, which also violates lepton number conservation (LNV).  Many quark flavour combinations are considered ($q\in[d,s,b]$ and $q^{\prime}\in[u,c]$, leading to $t[d,s,b][u,c]\ell$ vertices), together with a $W\rightarrow\ell\nu$ decay from the SM top quark, giving a dileptonic signature.  The $ee$, $e\mu$ and $\mu\mu$ channels are included, with one $b$-tagged jet and other kinematic requirements.  A single BDT is used to extract the BNV signal, and no significant excess over the background is found.  Limits are set separately for each BNV coupling, for $s$-channel or $t$-channel exchange, and branching fractions are extracted.
The upper limits on the branching fractions are illustrated in Figure~\ref{fig:BNV_HNL}(a) for this analysis.  They improve on the previous limits at $\sqrt{s}=8$~TeV~\cite{CMS:2013zol} by factors of $10^3$--$10^6$.

\section{Heavy Neutral Leptons}
\label{sec:HNL}

It is known that neutrinos must have mass, due to the presence of neutrino mixing.  A Type-I seesaw model~\cite{Yanagida} adds three heavy right-handed Majorana neutrinos $N_i$ ($i=1,3$), in addition to the SM neutrinos.  These $N_i$ couple to the SM charged leptons with strengths given by a mixing matrix $V$.  Many searches for heavy neutral leptons (HNL) have been carried out or are ongoing at the LHC.
The ATLAS analysis~\cite{ATLAS:2024fcs} discussed here searches for an HNL signature in a top quark decay from $t\bar{t}$ for the first time.  One top quark decays according to the SM, while the other includes a $t\rightarrow N\ell,~N\rightarrow W\ell$ decay chain; this leads to LNV of two units.  The model considered here includes only diagonal mixing terms, with a single $N$ coupling to one of $e$, $\mu$ or $\tau$ with strength $V_{\ell, N}$.  The final state includes two same-sign leptons ($ee$ or $\mu\mu$) and hadronic decays of two $W$ bosons. HNL signal samples are generated in the mass range 15--75 GeV and the HNL signal is distinguished from background using one BDT for low $m_N$ and another for high $m_N$.  Profile likelihood fits are applied to the $ee$ and $\mu\mu$ channels separately; the same data are reinterpreted for the $\tau\tau~(\rightarrow ee, \mu\mu)$ channels. Limits are set on the HNL cross-sections and coupling parameters, extending the ATLAS $ee$ and $\mu\mu$ limits in the range above $m_N = 50$~GeV\@.  Figure~\ref{fig:BNV_HNL}(b) shows the limits for the $\mu\mu$ channel.  

\section{Conclusion}

An extensive programme of top quark BSM searches is in progress with the ATLAS and CMS experiments at the Large Hadron Collider, including charged-lepton flavour violation, baryon number violation and lepton number violation via heavy neutral lepton production.  Typical branching ratio limits lie in the range of $10^{-6}$ to $10^{-8}$.  The Effective Field Theory approach is a useful tool for model-independent BSM searches, allowing limits to be set on Wilson coefficients for dimension-six four-fermion interactions.  As a further test of the SM, lepton flavour universality between electrons and muons has been tested very precisely via top processes: ${\cal B}(W\rightarrow\mu\nu)/{\cal B}(W\rightarrow e\nu)$ has been measured to $0.45\%$, the most precise value to date.  In the near future, the Run 3 datasets will allow further probes of the top quark interactions, using new techniques and with improved precision.







\end{document}